\documentclass[%
 reprint,
nofootinbib,
 amsmath,amssymb,
 aps,
]{revtex4-1}

\usepackage{graphicx}
\usepackage{dcolumn}
\usepackage{bm}
\usepackage{hyperref}
\hypersetup{
    colorlinks=true,
    linkcolor=blue,
    filecolor=blue,      
    urlcolor=blue,
    citecolor=blue
    }
\usepackage{amsmath,amssymb,amsfonts,bm}
\usepackage{ulem}
\usepackage[dvipsnames]{xcolor}

\begin{document}




\title{Constraining $\alpha$-cluster compactness in $^{16}\rm O$ and $^{20}\rm Ne$ at TeV energies using\\
 azimuthal anisotropy}




\author{Aswathy Menon Kavumpadikkal Radhakrishnan$^{1}$}\email[]{Aswathy.Menon@cern.ch}
\author{Suraj Prasad$^{2}$}\email[]{Suraj.Prasad@cern.ch}
\author{Neelkamal Mallick$^3$}\email[]{Neelkamal.Mallick@cern.ch}
\author{Raghunath Sahoo$^1$}\email[Corresponding author:]{Raghunath.Sahoo@cern.ch}
\author{Gergely Gábor Barnaföldi$^2$}\email[]{barnafoldi.gergely@wigner.hun-ren.hu}
\affiliation{$^1$Department of Physics, Indian Institute of Technology Indore, Simrol, Indore 453552, India}
\affiliation{$^2$HUN-REN Wigner Research Centre for Physics, 29-33 Konkoly-Thege Miklós Str., H-1121 Budapest, Hungary}
\affiliation{$^3$University of Jyv\"askyl\"a, Department of Physics, P.O. Box 35, FI-40014, Jyv\"askyl\"a, Finland}

\date{\today} 

\begin{abstract}
Anisotropic flow in ultra-relativistic light-ion collisions is sensitive to the initial geometry of the colliding nuclei. We investigate whether elliptic flow measurements can constrain the parameters of the proposed $\alpha$-clustered nuclear density distributions of $^{16}$O and $^{20}$Ne at LHC energies. Using the hybrid framework IP-Glasma+MUSIC+iSS+UrQMD, we simulate OO and Ne--Ne collisions at $\sqrt{s_{\mathrm{NN}}}=5.36$ TeV for the Woods--Saxon and $\alpha$-clustered configurations with varying cluster compactness. The elliptic flow coefficient $v_2\{2,|\Delta\eta|>1\}$ is calculated in the kinematic acceptances of ALICE, CMS, and ATLAS detectors and is compared with the Run~3 OO and Ne--Ne experimental measurements. It is observed that the final-state elliptic flow is significantly sensitive to the nuclear geometry, especially in OO collisions, where different configurations lead to distinct centrality dependencies and peak positions of $v_{2}$. By performing a systematic variation of the cluster size and inter-cluster separation in $^{16}$O and $^{20}$Ne nuclei, this work attempts to identify the cluster parameter range that provides the best agreement with the experimental data. These results show that the flow observables in TeV-energy light-ion collisions can be used to optimize the nuclear structure parameters of light nuclei.
\end{abstract}


\maketitle

\section{INTRODUCTION}
\label{sec:intro}

\begin{figure}[ht!]
\centering
\includegraphics[scale=0.38]{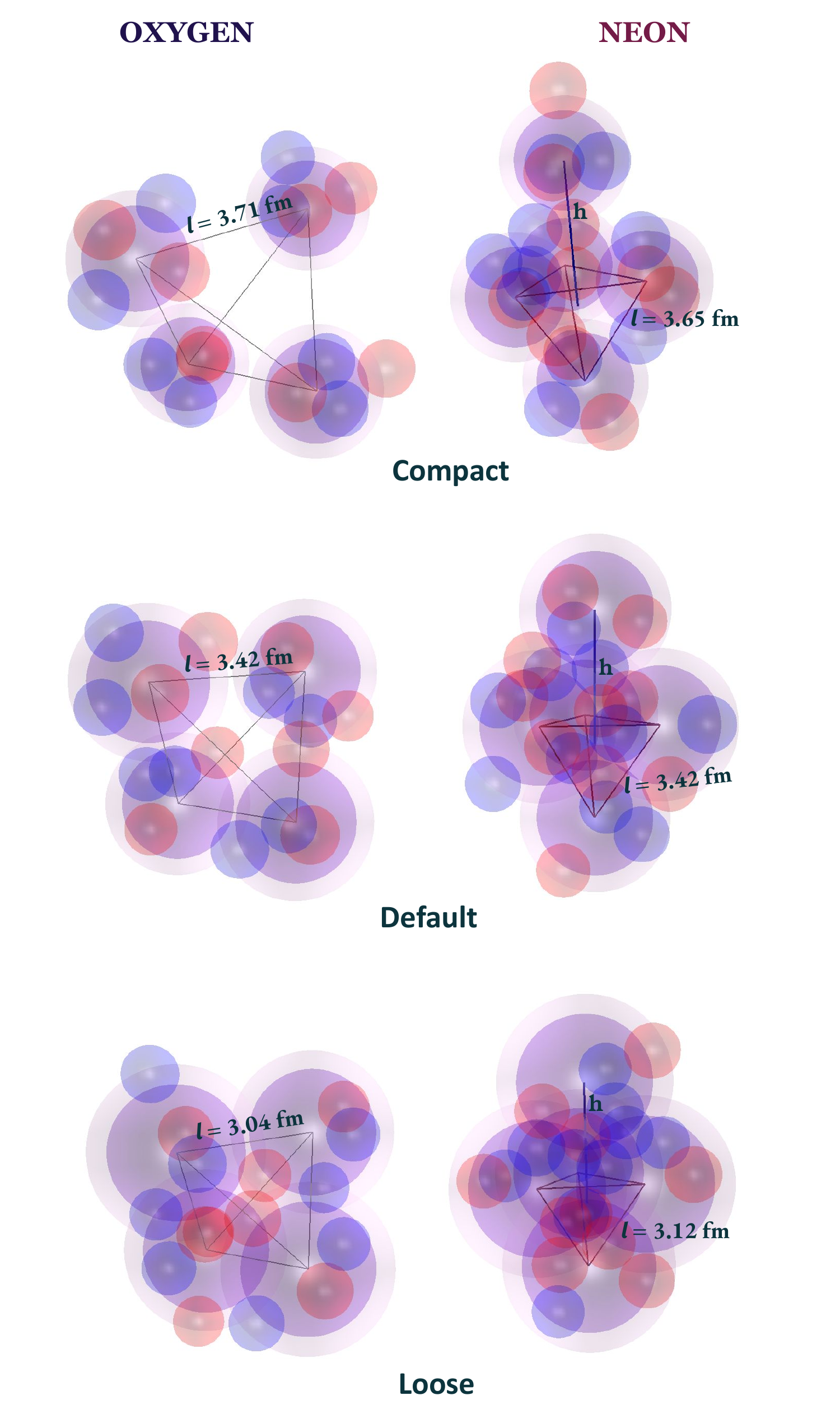}
\caption{Pictorial representation of compact, default and loose $\alpha$-clustered configurations of $^{16} \rm O$ (left column) and $^{20} \rm Ne$ (right column) nuclei. [Not to scale]}
\label{fig:ONe3config}
\end{figure}

Ultra-relativistic collisions of heavy ions are capable of recreating the extreme temperature and energy-density conditions that prevailed in the early Universe, resulting in the formation of Quark-Gluon Plasma (QGP)---a hot, dense soup of strongly interacting nuclear matter, consisting of deconfined quarks and gluons~\cite{Busza:2018rrf}. The collective expansion of QGP converts the initial geometric/spatial anisotropy into the observed momentum anisotropy of secondary particles produced, which is quantified by the flow harmonics, $v_{\rm n}$. Although \textit{pp} and \textit{pA} collisions have served as traditional baselines for QGP studies in heavy-ion systems, the observation of collective-flow-like signals in these collisions~\cite{CMS:2010ifv, ALICE:2013snk, CMS:2015fgy, CMS:2016fnw, ALICE:2024vzv} has stimulated considerable interest in the possibility of QGP droplet formation in progressively smaller systems~\cite{Constantin:2025ova, CMS:2026qef, ALICE:2026zck, ATLAS:2026fos, ATLAS:2026zgq}. 

One of the major limitations in comprehending collective flow signatures in hadronic collisions is the uncertainty associated with their initial conditions, particularly due to the significant role played by the proton sub-structure~\cite{ALICE:2025luc}. Symmetric light-ion collisions, on the other hand, offer a more controlled environment where the initial spatial anisotropy is more strongly influenced by the nucleonic arrangement within the colliding nuclei~\cite{Giacalone:2024luz}. In this context, the recent light-ion programs at the Large Hadron Collider (LHC), such as oxygen--oxygen (OO) and neon--neon (Ne--Ne) collisions, provide a unique opportunity to investigate the onset of QGP-like phenomena in systems of intermediate size~\cite{ALICE:2025luc, ATLAS:2025nnt, ATLAS:2026fos, CMS:2025tga, CMS:2026qef, ATLAS:2026zgq, ALICE:2026zck, Huss:2020dwe}. Moreover, investigating collectivity using $^{16}\rm O$ and $^{20}\rm Ne$ nuclei has an additional benefit, i.e., both nuclei have comparable mass numbers, and therefore, OO and Ne--Ne collisions undergo similar bulk evolution and should exhibit similar final-state collective responses. Any differences observed in their final-state flow coefficients can be mapped to the frozen imprints of the structural differences in the intrinsic nuclear configurations at the instant of these relativistic collisions~\cite{Giacalone:2024luz, ALICE:2025luc}. Low-energy nuclear experiments and nuclear structure calculations suggest that the $^{16}\mathrm{O}$ nucleus may exhibit a tetrahedral four-$\alpha$ cluster configuration~\cite{Bijker:2014tka, Wang:2019dpl, He:2014iqa, He:2021uko, Otsuka:2022bcf}, whereas $^{20}\mathrm{Ne}$ is predicted to possess a characteristic ``bowling-pin''-like intrinsic geometry, consistent with a $^{16}\mathrm{O}+\alpha$ cluster arrangement~\cite{Giacalone:2024luz, Bijker:2020gbl}. These intrinsic cluster structures are associated with characteristic nuclear deformations, giving rise to predominant octupole deformation in $^{16}\mathrm{O}$ and quadrupole deformation in $^{20}\mathrm{Ne}$~\cite{Wang:2024ulq}. Such differences in the initial nuclear geometry are expected to influence the spatial anisotropies of the collision system and, consequently, the development of collective flow in the final state~\cite{ALICE:2018lao}. In particular, the quadrupole deformation of $^{20}\mathrm{Ne}$ is anticipated to leave a distinctly different imprint on the final-state azimuthal anisotropy coefficients ($v_{n}$) compared to the octupole deformed configurations of $^{16}\mathrm{O}$~\cite{Giacalone:2024luz}. Therefore, measurements of $v_n$ in OO and Ne--Ne collisions can provide valuable information not only on collective flow dynamics but also on the degree of nuclear deformation and possible $\alpha$-clustering in these light nuclei, especially in ultra-central collisions where geometric effects are expected to be enhanced~\cite{ALICE:2025luc,
Wang:2024ulq, Giacalone:2024luz}.

The possibility of probing nuclear structure through relativistic collisions of light nuclei has attracted considerable attention in recent years. Hydrodynamic and transport-model studies have demonstrated that initial-state geometric features arising from $\alpha$-cluster configurations can survive the medium evolution and manifest themselves in final-state flow observables~\cite{Constantin:2025ova, Ding:2023ibq, MenonKavumpadikkalRadhakrishnan:2025apq, Behera:2023nwj, Behera:2021zhi, R:2024eni}, their fluctuations~\cite{Prasad:2024ahm}, symmetric cumulants~\cite{Shafi:2025feq}, and in the correlation of different symmetry plane angles~\cite{Prasad:2026osg}. In particular, OO collisions have been proposed as a sensitive probe of tetrahedral four-$\alpha$ clustering in $^{16}\mathrm{O}$, with measurable effects predicted in observables which are highly influenced by event-by-event density fluctuations in the collision overlap regions, such as $v_3$ and fluctuations in $v_2$~\cite{YuanyuanWang:2024sgp}. Similarly, recent theoretical investigations of Ne--Ne collisions suggest that the intrinsic quadrupole deformation and possible cluster structure of $^{20}\mathrm{Ne}$ can significantly modify the initial eccentricities, leading to enhanced elliptic flow in ultra-central events~\cite{Giacalone:2024luz, Li:2025hae}. Recent experimental measurements of anisotropic flow coefficients provide reasonable support for these predictions, showing trends that are consistent with the expected effects of nuclear deformation and clustering on the final-state collective behavior~\cite{ALICE:2025luc, ATLAS:2025nnt, CMS:2025tga}.

While existing studies have primarily focused on ultra-central collisions, investigating the centrality dependence of the observed effects is equally important for understanding the role of nuclear geometry. In experiments, collision centrality is typically estimated using the produced charged-particle multiplicity, which serves as a reasonable proxy for the collision impact parameter in heavy-ion collisions. However, as the size of the colliding nuclei decreases, the correlation between impact parameter and charged-particle multiplicity becomes weaker. Similar limitations are also observed among different multiplicity-based centrality estimators used within and across experiments~\cite{Prasad:2025yfj}. As a result, an experimentally selected centrality class may contain events originating from a broad range of true impact parameters. Consequently, the measured ``central'' events may not correspond to genuinely central collisions. It is therefore important to investigate the effects associated with the true collision geometry, characterized by the impact parameter, and to identify possible signatures of these effects in the final-state observables. This is particularly relevant for light-ion collisions, where centrality fluctuations and event-selection biases can dilute geometric effects related to nuclear deformation and clustering~\cite{Schukraft:2012ah}. For example, in our previous study, a peak-like structure was observed in the elliptic flow coefficient for clustered nuclear configurations when the analysis was performed using impact parameter-based centrality selections, whereas no such feature was found for the corresponding Woods--Saxon configurations~\cite{MenonKavumpadikkalRadhakrishnan:2025apq}. However, this peak-like structure becomes significantly weaker when multiplicity-based centrality estimators are employed~\cite{Prasad:2024ahm}. Interestingly, similar features have also been reported in experimental measurements by the ALICE, ATLAS, and CMS collaborations~\cite{ALICE:2025luc, ATLAS:2025nnt, CMS:2025tga}. These observations suggest that the interplay between the intrinsic nuclear geometry and the method used to define collision centrality may play an important role in the origin of such effects. A systematic investigation of geometry-based centrality selections along with various cluster profile implementations in $^{16}$O and $^{20}$Ne nuclei may therefore provide valuable insights into the influence of nuclear deformation and clustering on the final-state observables.

In this Letter, we aim to investigate the effects of the initial nuclear configurations of the colliding $^{16}\rm O$ and $^{20}\rm Ne$ nuclei on the final-state anisotropic flow coefficients within a hybrid model framework. Elliptic flow coefficient $v_2\{2,|\Delta\eta|>1\}$ is separately computed in the ALICE, CMS, and ATLAS kinematic acceptances and is compared to the available experimental data, wherever applicable. The results are presented for four different configurations, for both oxygen and neon nuclei, namely, the Woods-Saxon, compact $\alpha$-cluster, default $\alpha$-cluster, and loose $\alpha$-cluster.

\section{Nuclear Density Profiles} 
\label{sec:nuclearprofiles}
\begin{table*}[ht!]
\centering
\caption{Woods-Saxon and $\alpha$-cluster configurations for $^{16} \rm O$ nucleus and $^{20} \rm Ne$ nucleus~\cite{deVries:1987}.}
\begin{tabular}{ccccccc}
\hline
\textbf{Nuclear Distribution}  & \multicolumn{3}{|c}{\textbf{OXYGEN} ($r_{\rm rms}=$ 2.68~fm)} & \multicolumn{3}{|c}{\textbf{NEON} ($r_{\rm rms}=$ 3.07~fm)} \\ 
\hline 
\hline
\textbf{Woods-Saxon}  & $r_{0}$ (fm) & $w$ & $a$ & $r_{0}$ (fm) & $w$ & $a$ \\

 &  2.608    &  -0.051    &  0.513  & 2.791   &  -0.168  & 0.698  \\ \hline
 
\textbf{$\alpha$-cluster configurations:}   & $r_{\alpha}$  (fm) & $l$ (fm) & $r_{\alpha}/l$ & $r_{\alpha}$ (fm) & $l$  (fm) & $r_{\alpha}/l$ \\ \hline
\textbf{Default} &  1.676 & 3.42  & 0.49 &  1.676 & 3.42 & 0.49\\ 
\textbf{Compact} & 1.420  & 3.71  & 0.38 & 1.386  & 3.65  & 0.38 \\ 
(Comparison to default) & (84.7\%) & (108.5\%) & (77.6\%) & (82.7\%) & (106.7\%) & (77.6\%)  \\
\textbf{Loose} & 1.930  & 3.04  & 0.635 & 1.982 & 3.12  & 0.635 \\ 
(Comparison to default) & (115.2\%) & (88.9\%) & (129.6\%)  & (118.3\%) & (91.2\%) & (129.6\%) \\
\hline 

\end{tabular}
\label{tab:ConfigTable}
\end{table*}

 \textit{Woods-Saxon:} The positions of nucleons inside $^{16}\rm O$ and $^{20}\rm Ne$ nuclei are randomly sampled using the following Woods-Saxon density profile in terms of a three-parameter Fermi (3pF) distribution~\cite{Loizides:2025ule}:
\begin{equation}
\rho(r) = \frac{\rho_{0} \Big(1+ w \big(\frac{r}{r_{0}}\big)^{2}\Big)}{1 + \exp\big(\frac{r - r_{0}}{a}\big)}, 
\label{eq:WS}
\end{equation}
where $\rho_{0}$ and $\rho(\rm r)$ are the nuclear charge densities at the radial distance $r=0$ and at $``r"$ respectively. The mean nuclear radius ($r_0$), nuclear deformation parameter ($w$), and skin depth ($a$) used for both nuclei are tabulated in Table~\ref{tab:ConfigTable}. \\
 \textit{Tetrahedral $\alpha$-cluster:} The four $\alpha$-clusters in $^{16}\rm O$ nucleus are placed at the four vertices of a regular tetrahedron with inter-cluster distance (sidelength) $l$, while in $^{20}\rm Ne$ nucleus, the fifth $\alpha$-cluster is placed along the symmetry axis above the tetrahedral centroid, at distance $h = 1.15\times l$, creating a bowling-pin geometry~\cite{YuanyuanWang:2024sgp, Li:2025hae}. Inside each $\alpha$-cluster with the root-mean-squared radius $r_{\alpha}$, the four nucleon positions are sampled independently from a Gaussian radial distribution as:
\begin{equation}
    P(r) \propto r^{2} \exp{\Big(-\frac{r^{2}}{2\sigma^{2}}\Big)}
\end{equation}
with $\sigma=r_{\alpha}/\sqrt{3}$, where a minimum separation constraint of $d_{\rm min}=0.4$~fm is maintained between any two nucleon pair within the $\alpha$-cluster~\cite{Loizides:2017ack}.

To study the effects due to the compactness of the $\alpha$-clustering in the colliding oxygen and neon nuclei, we perform a systematic variation of $r_{\alpha}$ and $l$ such that the root-mean-squared radius ($R_{\rm rms}$) of the respective nucleus (estimated as shown below) is not altered~\cite{YuanyuanWang:2024sgp}.
\begin{eqnarray}
   R_{\rm rms_{Ox}} = \sqrt{\langle r_{\rm Ox}^{2}\rangle} \equiv \sqrt{\frac{3}{8}l^{2} + r_{\alpha}^{2}} =  2.68~\rm fm  \nonumber\\ 
     R_{\rm rms_{Ne}} = \sqrt{\langle r_{\rm Ne}^{2}\rangle} \equiv \sqrt{0.5645l^{2} + r_{\alpha}^{2}} = 3.07~\rm fm 
\end{eqnarray}
A compact $\alpha$-clustered nucleus refers to that with smaller $r_{\alpha}$ and longer $l$, while a loose one is where a shorter $l$ and larger $r_{\alpha}$ give rise to a configuration where the cluster boundaries are nearly overlapping, as depicted in Fig.~\ref{fig:ONe3config}. For easy comparison, the $r_{\alpha}$ and $l$ values are chosen such that the $r_{\alpha}/l$ ratios for compact, default, and loose $\alpha$-cluster configurations are the same for both $^{16}\rm O$ and $^{20}\rm Ne$ nuclei, as can be found in Table~\ref{tab:ConfigTable}.\\

\section{Model description} 
\label{sec:model}

In this work, the $^{16}\rm O$--$^{16}\rm O$ and $^{20}\rm Ne$--$^{20}\rm Ne$ (hereafter referred to as $\rm O\rm O$ and $\rm Ne$--$\rm Ne$ respectively) collisions are simulated at $\sqrt{s_{\rm NN}} = 5.36$~TeV using the hybrid hydrodynamic model, IP-Glasma+MUSIC+iSS+UrQMD as in our previous work~\cite{MenonKavumpadikkalRadhakrishnan:2025apq}. In this framework, the event-by-event initial conditions from IP-Glasma at $\tau_{\rm switch}~=~0.4$~fm are evolved using viscous hydrodynamics in MUSIC within a (2+1)D boost-invariant framework (EoS \texttt{s95p-v1.2}, $\eta/s = 0.12$). When the local energy density drops to $\varepsilon_{\rm switch}=0.18~\rm GeV/fm^{3}$, a freeze-out hypersurface is constructed and the particlization is performed using \texttt{iSS(iSpectraSampler)} via the Cooper-Frye prescription. To improve statistical precision without rerunning the hydrodynamical evolution, each IP-Glasma+MUSIC event is oversampled 200 times using \texttt{iSS}. The sampled hadrons are subsequently propagated through the microscopic transport model, UrQMD (Ultra-relativistic Quantum Molecular Dynamics, version 3.4, default settings), to account for hadronic rescatterings and decays, resulting in a more realistic dynamical freeze-out of different hadronic species~\cite{Schenke:2020mbo,
MenonKavumpadikkalRadhakrishnan:2025apq}.

\section{Flow coefficient estimation} 
\label{sec:2pCumu}

Anisotropic flow coefficients, which characterize collectivity, are quantified using the coefficients of the Fourier expansion of the azimuthal distribution of the particles in the final state, given as follows~\cite{Voloshin:1994mz},
\begin{equation}
\frac{\mathrm{d}N}{\mathrm{d}\phi}\propto 1+\sum_{n=1}^{\infty}2v_{n}\cos[n(\phi-\psi_{n})] \ 
\label{eq:fourierexpansion}
\end{equation}
where, $\phi$ and $\psi_n$ are the azimuthal angle and $n$\textsuperscript{th} harmonic symmetry plane angle respectively, while $v_n$ denotes the anisotropic flow coefficients which can be calculated as $v_{n}=\langle\cos[n(\phi-\psi_{n})]\rangle$. Since $\psi_n$ cannot be measured directly, this work resorts to the two-particle Q-cumulant method to estimate the flow coefficients $v_{2}$ (elliptic flow) and $v_{3}$ (triangular flow)~\cite{Bilandzic:2010jr}. Combining the 200 UrQMD events oversampled from the same hydrodynamic freeze-out surface into a ``super-event"(SE), we define the $n$\textsuperscript{th} order flow vector as~\cite{McDonald:2016vlt},
\begin{equation}
    Q^{\rm SE}_{n}=\sum_{j=1}^{N_{\rm sample}}\sum_{k=1}^{M_{j}}\mathrm{e}^{in\phi_{jk}}
    \label{eq:QnSE} \ .
\end{equation}
where $M_{j}$ is the multiplicity of the $j$\textsuperscript{th} hadronic sample of the SE, and $\phi_{jk}$ is the azimuthal angle of the $k$\textsuperscript{th} particle in this sample. 
Non-flow contributions are suppressed by using super-sub-events, $A$ and $B$, with multiplicities $M_A$ and $M_B$ respectively, separated by $|\Delta\eta|>1$, giving the two-particle correlation by~\cite{Zhou:2015iba},
\begin{equation}
    \langle 2 \rangle _{\Delta \eta} = \frac{Q_{n}^{A} \cdot Q_{n}^{B *}} {M_{A} \cdot M_{B}} \ .
\end{equation}
The corresponding two-particle Q-cumulant is given by $c_{n}\{2, |\Delta\eta|\} = \langle \langle 2 \rangle \rangle _{\Delta\eta}$,
where $\langle\langle\dots\rangle\rangle$ denotes the average over the super events and the superscript `*' represents the complex conjugate. Finally, the anisotropic flow coefficients are calculated by $v_{n}\{2, |\Delta\eta|\} =\sqrt{c_{n}\{2, |\Delta\eta|\}}$.
%
%
%
%
\begin{figure*}[ht!]
\centering
\includegraphics[scale=0.29]{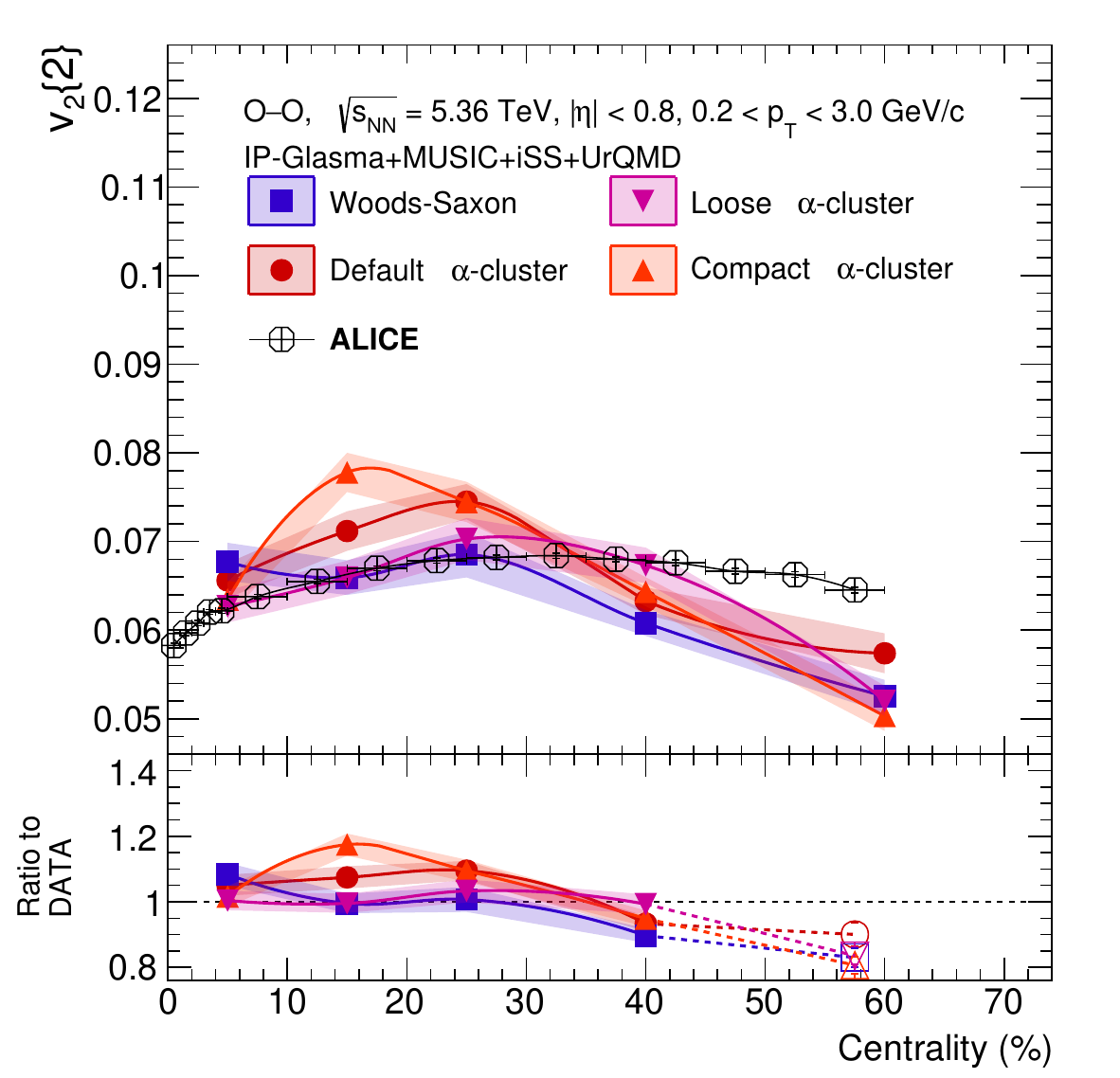}
\includegraphics[scale=0.29]{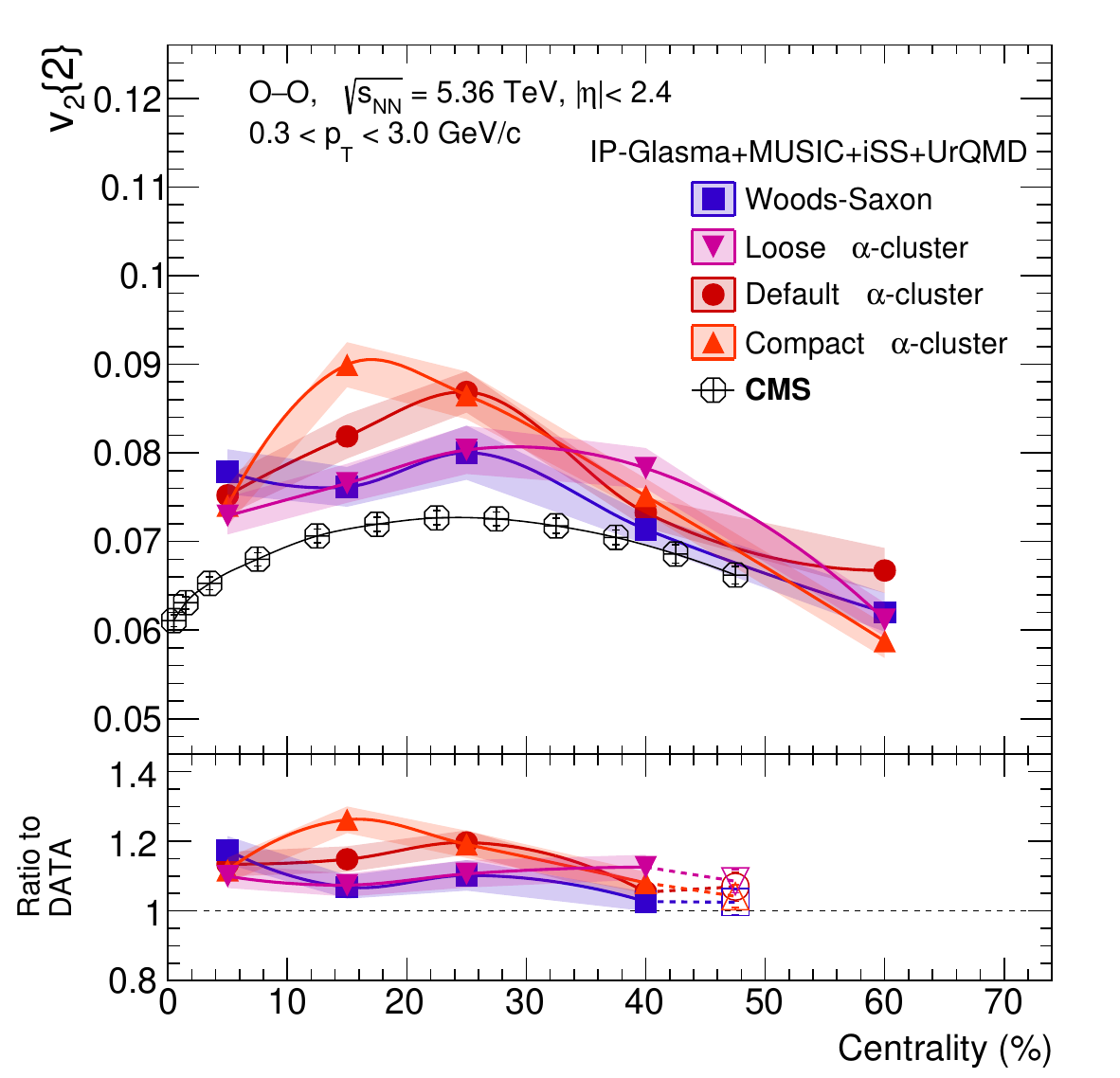}
\includegraphics[scale=0.29]{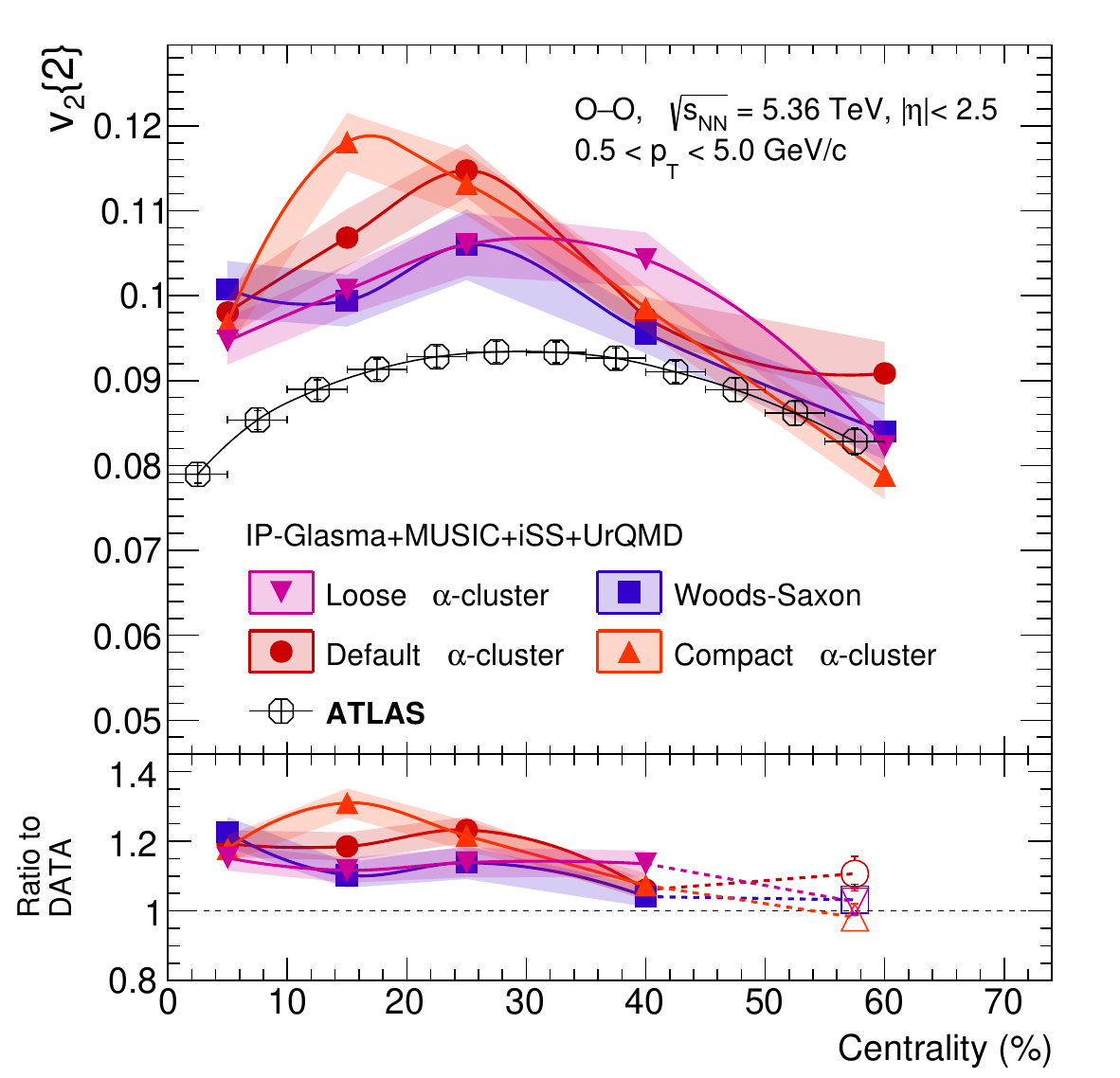}
\includegraphics[scale=0.29]{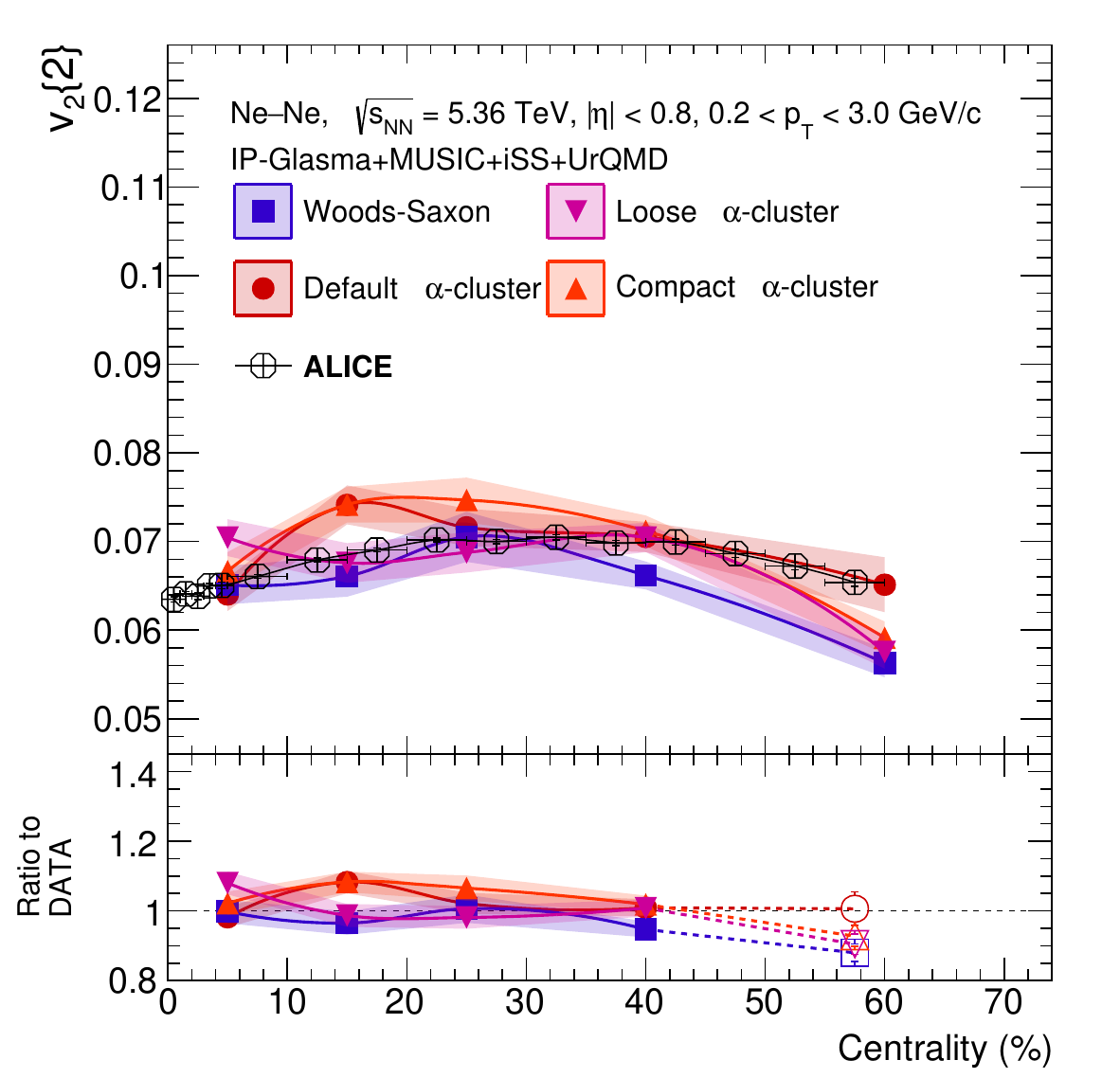}
\includegraphics[scale=0.29]{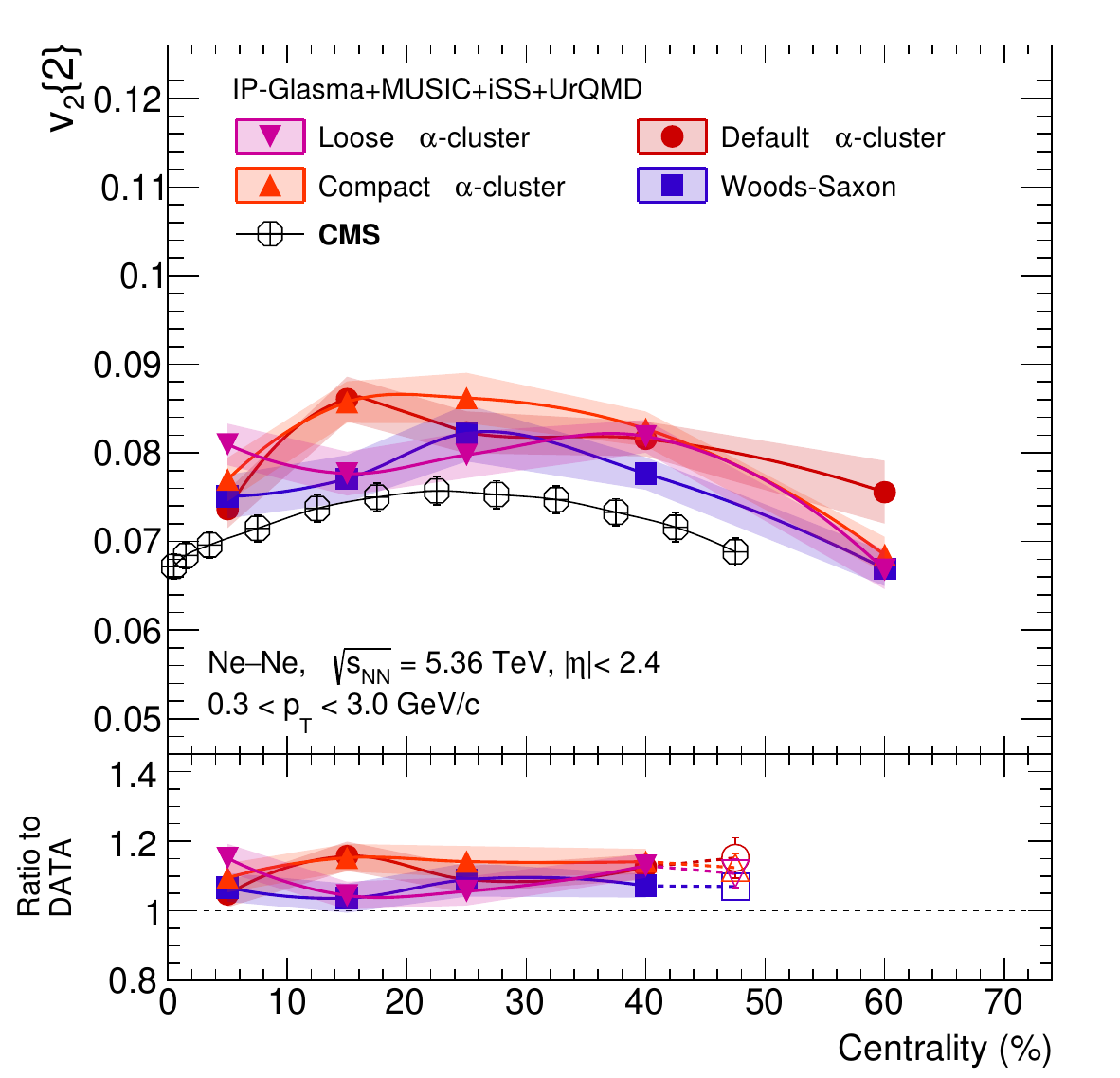}
\includegraphics[scale=0.29]{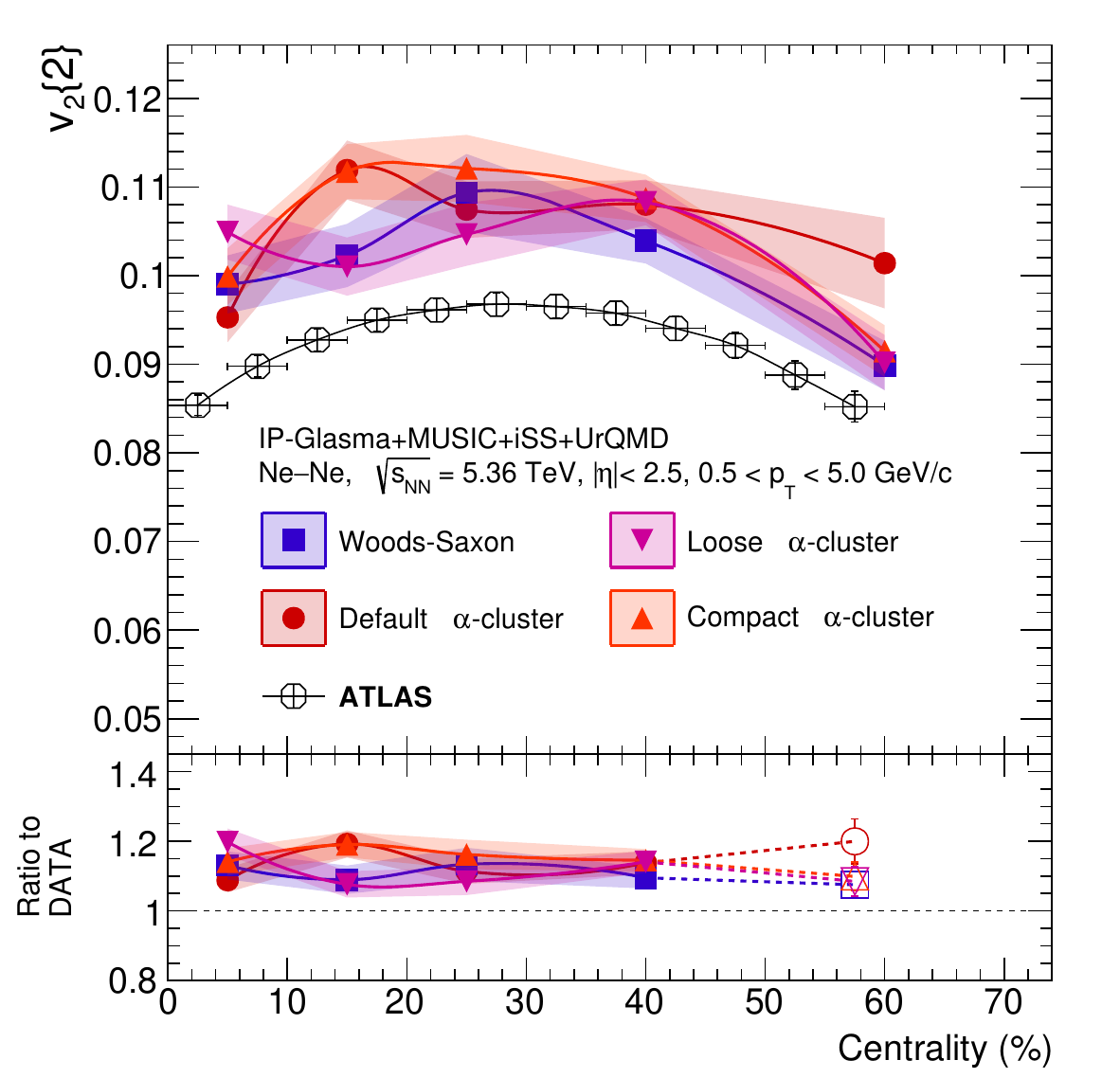}
\caption{$v_{2}\{2\}$ versus centrality for OO (upper row) and Ne--Ne (lower row) collisions at $\sqrt{s_{\rm NN}}~=~5.36$~TeV calculated for ALICE, CMS, and ATLAS kinematic regimes (left, middle, right columns respectively) using IP-Glasma+MUSIC+iSS+UrQMD model compared with data~\cite{ALICE:2025luc, ATLAS:2025nnt, CMS:2025tga}. Error bands incorporate the statistical uncertainties from model estimations.} 
\label{fig:NeO-v23-3expmnts}
\end{figure*}
%

\section{Results} 
\label{sec:R&D}
Elliptic flow, $v_{2}$, reflects the interplay between the initial eccentricity ($\epsilon_{2}$) of the collision system and the hydrodynamic response of the created medium, which converts spatial anisotropy into the final-state momentum anisotropy. The initial anisotropy is, in turn, governed by the intrinsic configuration of the colliding nuclei, sub-nucleonic and nucleonic fluctuations, as well as the shape of the collision overlap region. Since the impact parameter ($b$) is a quantity that geometrically characterizes the initial overlap region of the colliding nuclei, it is therefore most suitable for investigating the role of initial nuclear geometry in shaping final-state flow observables, as far as model-based studies are concerned. Hence, throughout this work, an impact parameter-based centrality selection is performed on OO and Ne--Ne events with the centrality class intervals being (0--10)$\%$, (10--20)$\%$, (20--30)$\%$, (30--50)$\%$ and (50--70)$\%$. 

Figure~\ref{fig:NeO-v23-3expmnts} shows the $v_{2}$ estimated from the two-particle cumulant method as a function of centrality, for OO (upper row) and Ne--Ne (lower row) collisions at $\sqrt{s_{\rm NN}}~=~5.36$~TeV for the four nuclear density profiles under study, using the IP-Glasma+MUSIC+iSS+UrQMD model. The left, middle, and right columns in Fig.~\ref{fig:NeO-v23-3expmnts} correspond to flow calculations performed using the kinematic regimes of ALICE ($|\eta|<0.8$, $0.2<p_{\rm T}<3.0$~GeV/$c$), CMS ($|\eta|<2.4$, $0.3<p_{\rm T}<3.0$~GeV/$c$), and ATLAS ($|\eta|<2.5$, $0.5<p_{\rm T}<5.0$~GeV/$c$) experiments, respectively. The bottom panel shows the ratio of $v_{2}$ from each of the nuclear profiles to the respective experimental data points~\footnote{The first four points in the ratio panels are the interpolation of data to match the model-centrality classes, while the fifth points represented by hollow markers use interpolated model prediction corresponding to the last available data point.}. As visible in the left column of Fig.~\ref{fig:NeO-v23-3expmnts}, the characteristic rise-peak-fall trend for $v_{2}\{2\}$ with respect to collision centrality observed in experimental results is also exhibited by the different nuclear profiles in our model-based study, but in varying degrees. In the most central OO collisions, $v_{2}\{2\}$ starts from similar values for compact, loose, and default cases of $\alpha$-clustering to increase up to mid-central collisions. However, the rate of rise of $v_{2}\{2\}$ is higher for the compact-cluster case than the default-cluster case, while the loose-cluster case has the lowest slope among the three $\alpha$-clustering configurations. It is worth noting that $v_{2}\{2\}$ from the compact-cluster configuration attains its peak value in the (10--20$\%$) centrality range, while all the other configurations--- the Woods-Saxon, loose, and default $\alpha$-clustering--- peak in the (20--30$\%$) centrality class, with the Woods-Saxon nuclear profile exhibiting the least prominent peak. In comparison to the rest of the nuclear profiles, the $v_{2}\{2\}$ from the loose $\alpha$-cluster case exhibits the best overlap with the ALICE data points of OO collisions at $\sqrt{s_{\rm NN}}=5.36$~TeV in the (0--40$\%$) centrality range. 

In Fig.~\ref{fig:NeO-v23-3expmnts}, for Ne--Ne collisions, it is observed that the sensitivity of $v_{2}\{2\}$ to nuclear profiles exists but is not as distinct as for OO collisions. Unlike in the case of OO collisions, peak positions of $v_{2}\{2\}$ are attained at (10--20$\%$) centrality class by the default $\alpha$-cluster, while the $v_{2}\{2\}$ peaks in the (20--30$\%$) class for the Woods-Saxon configuration. In contrast to OO collisions, there is no sharp $v_{2}\{2\}$ peak exhibited by the compact $\alpha$-cluster in Ne--Ne collisions; instead, it behaves like a plateau in the (10--30$\%$) centrality range. The loose $\alpha$-clustered Ne--Ne collisions show a fall-rise-fall trend for $v_{2}\{2\}$ where the maximum value of $v_{2}\{2\}$ is obtained in the most central and in the (30--50$\%$) centrality collisions. This time, $v_{2}\{2\}$ from the Woods-Saxon distribution has a fair similarity with ALICE data points in the (0--30$\%$) centrality range, while in the (30--70$\%$) region, the default $\alpha$-clustered trend is found to match the data curve efficiently. Turning to the CMS and ATLAS acceptances, we see that $v_{2}\{2\}$ from all nuclear profiles trend the same as in ALICE kinematic cuts, except for the fact that the magnitudes of $v_{2}\{2\}$ increase as one moves from ALICE to CMS to ATLAS, due to the increasing particle acceptance range in pseudorapidity and $p_{\rm T}$. Moreover, compared to the other two kinematic regimes studied, the $v_{2}\{2\}$ magnitudes from the model are in better agreement with the ALICE data points than with those of CMS or ATLAS. This is because the study uses MUSIC hydrodynamics in a 2+1D configuration that does not account for evolution along the longitudinal direction; hence, it is not expected to describe the experimental data over broader pseudorapidity ranges.

\begin{figure*}
\centering
\includegraphics[scale=0.42]{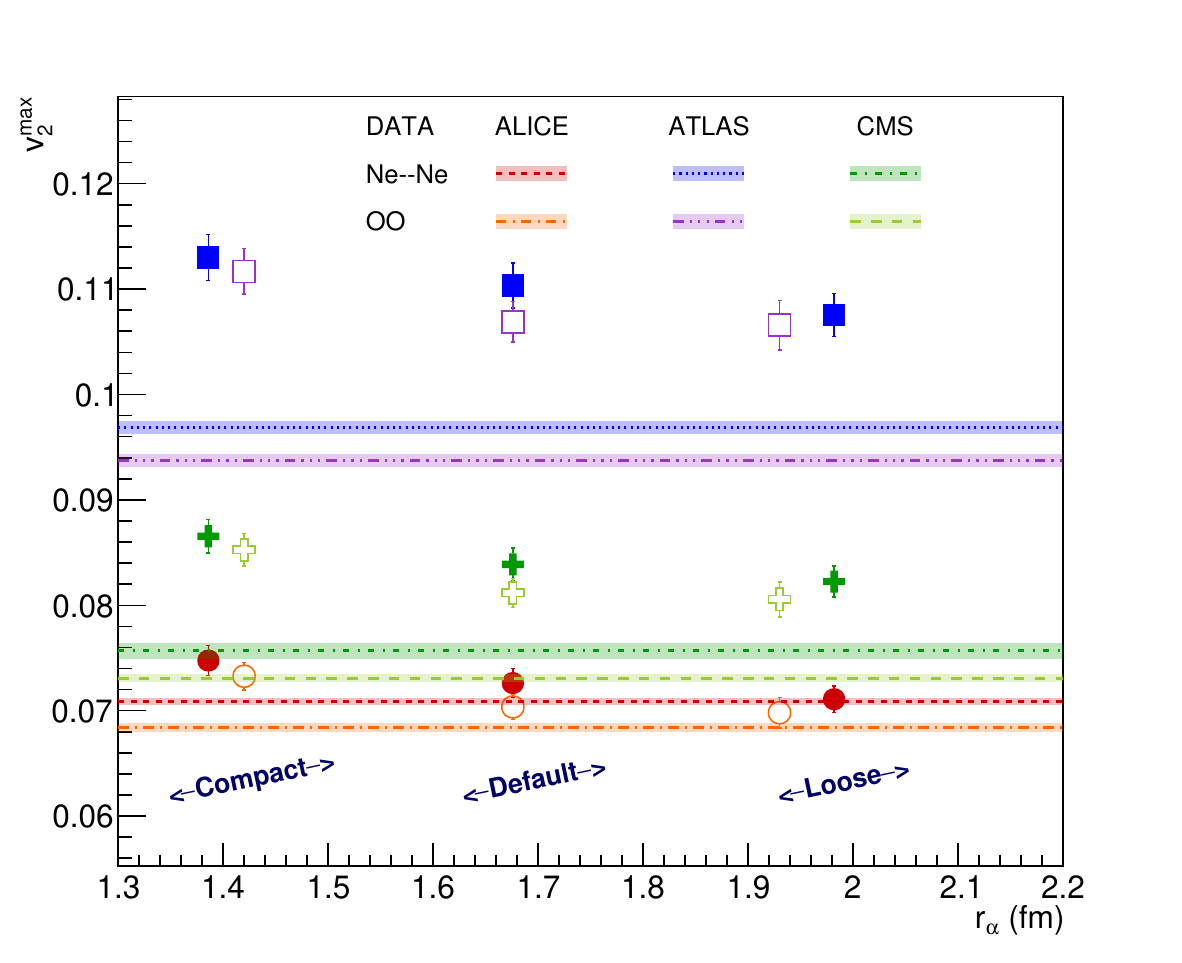}
\includegraphics[scale=0.42]{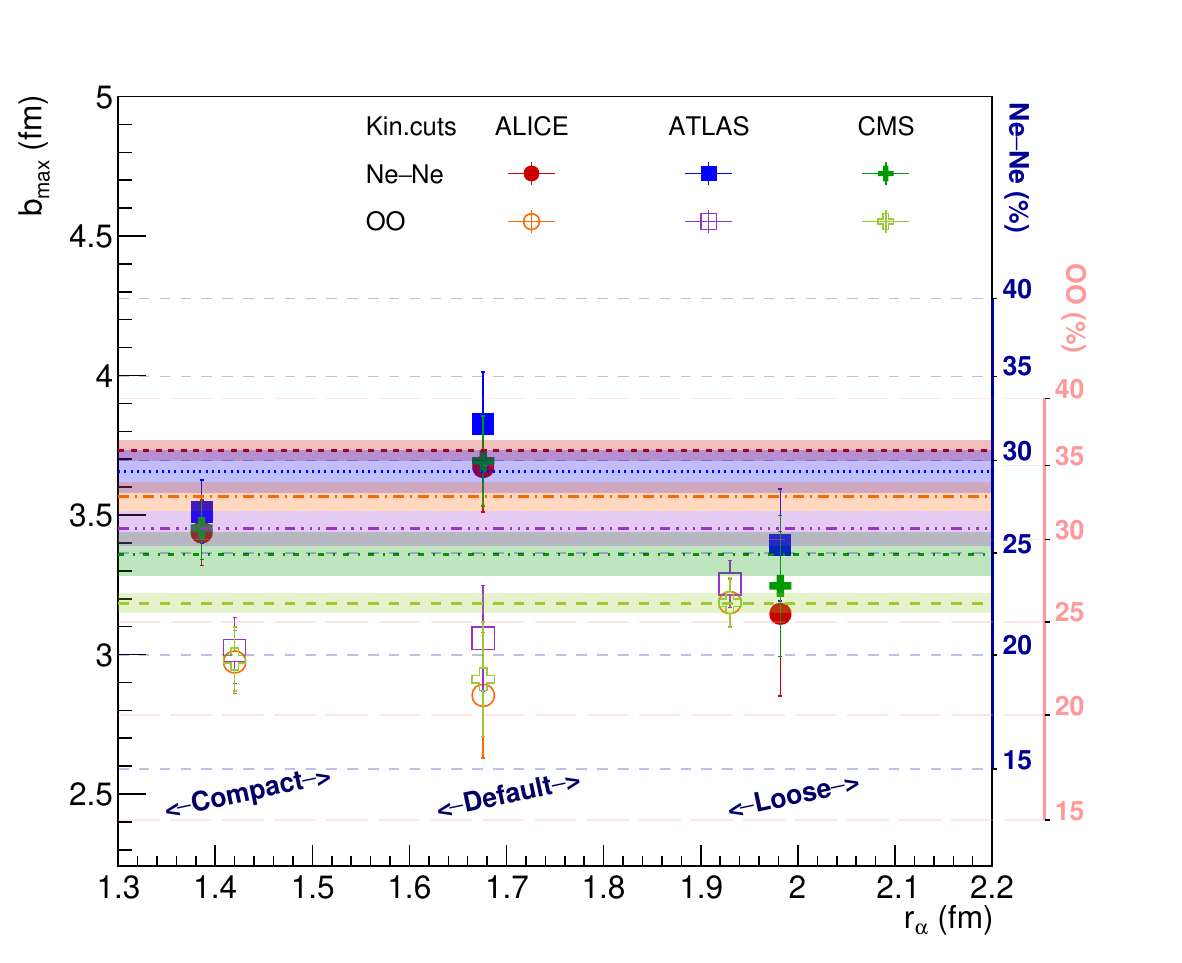}
\includegraphics[scale=0.42]{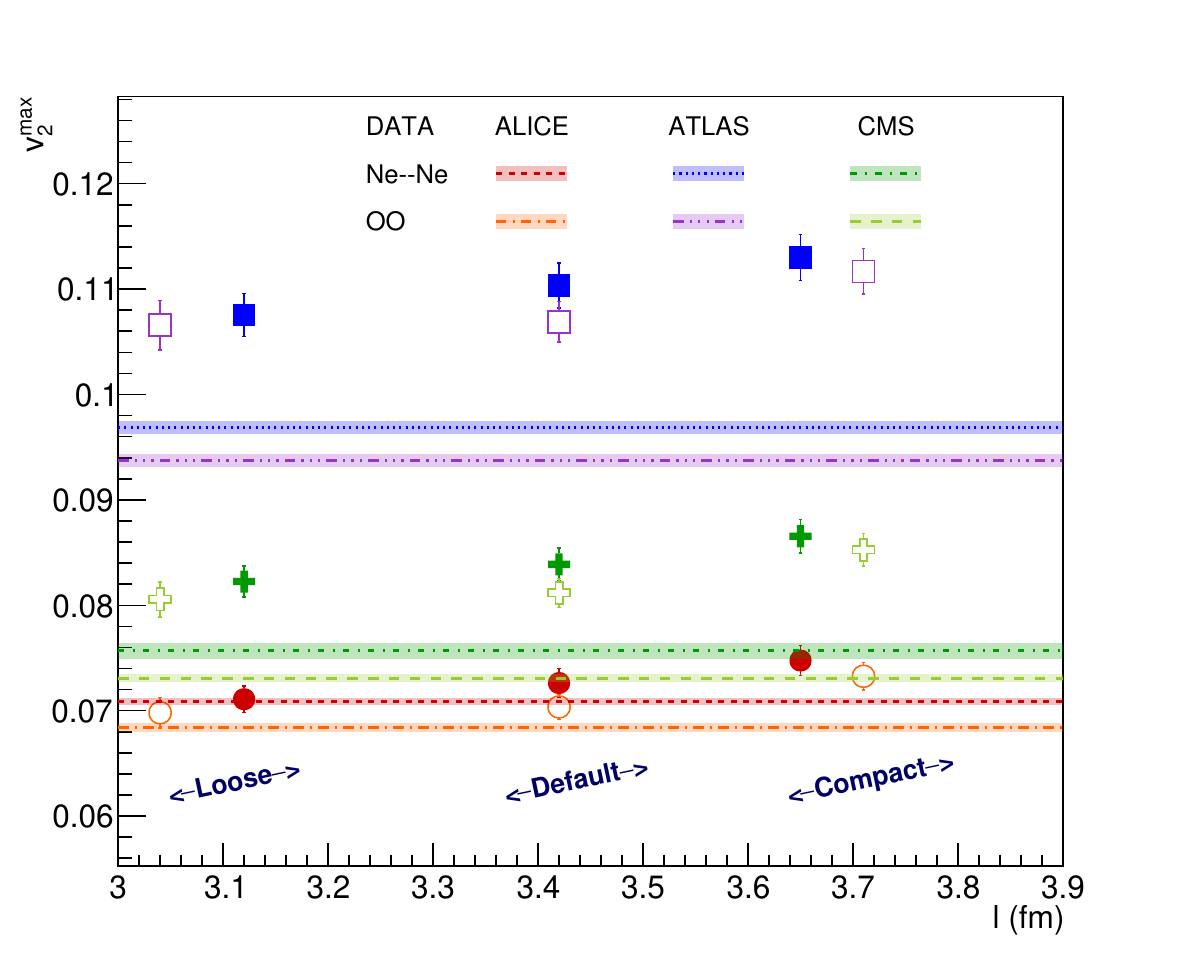}
\includegraphics[scale=0.42]{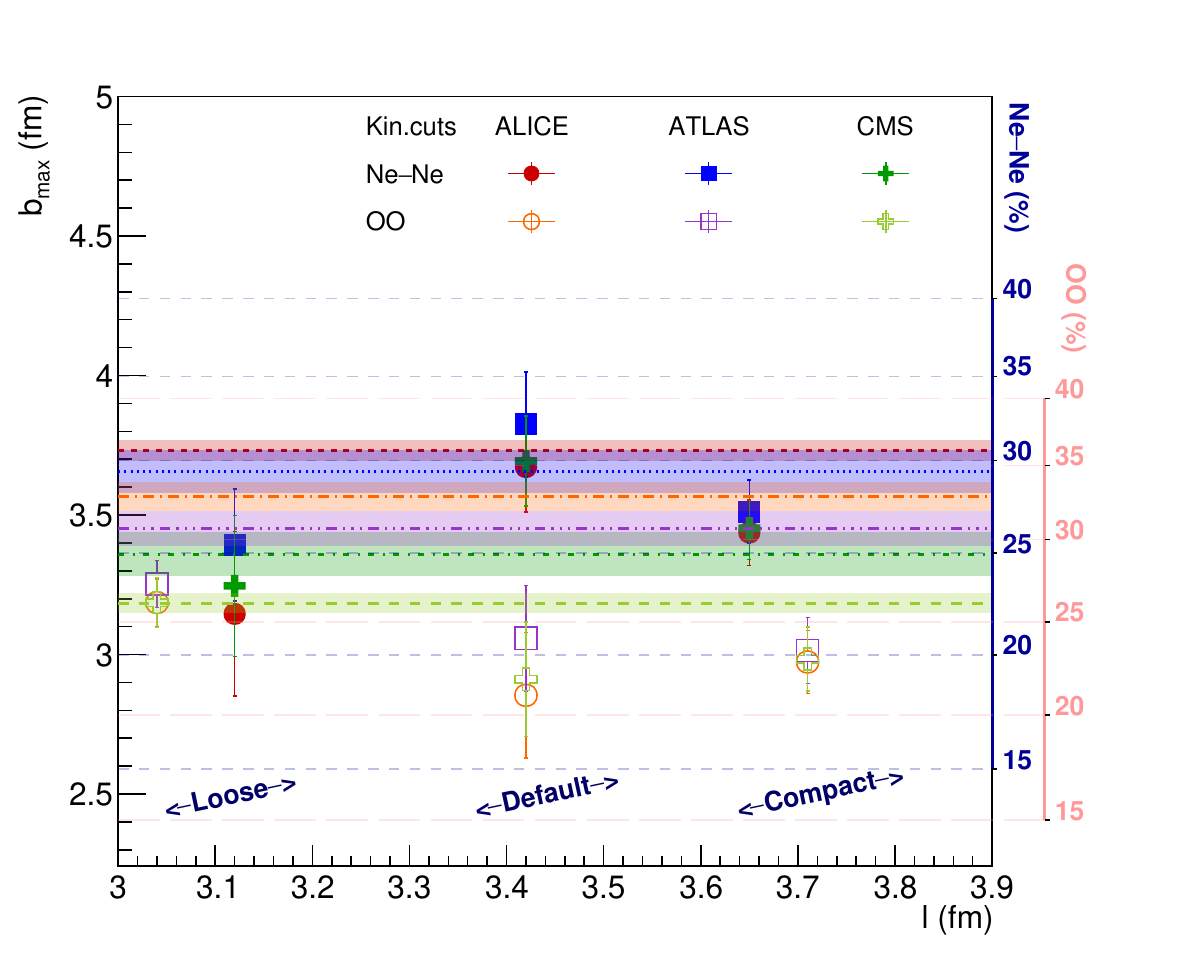}
\includegraphics[scale=0.42]{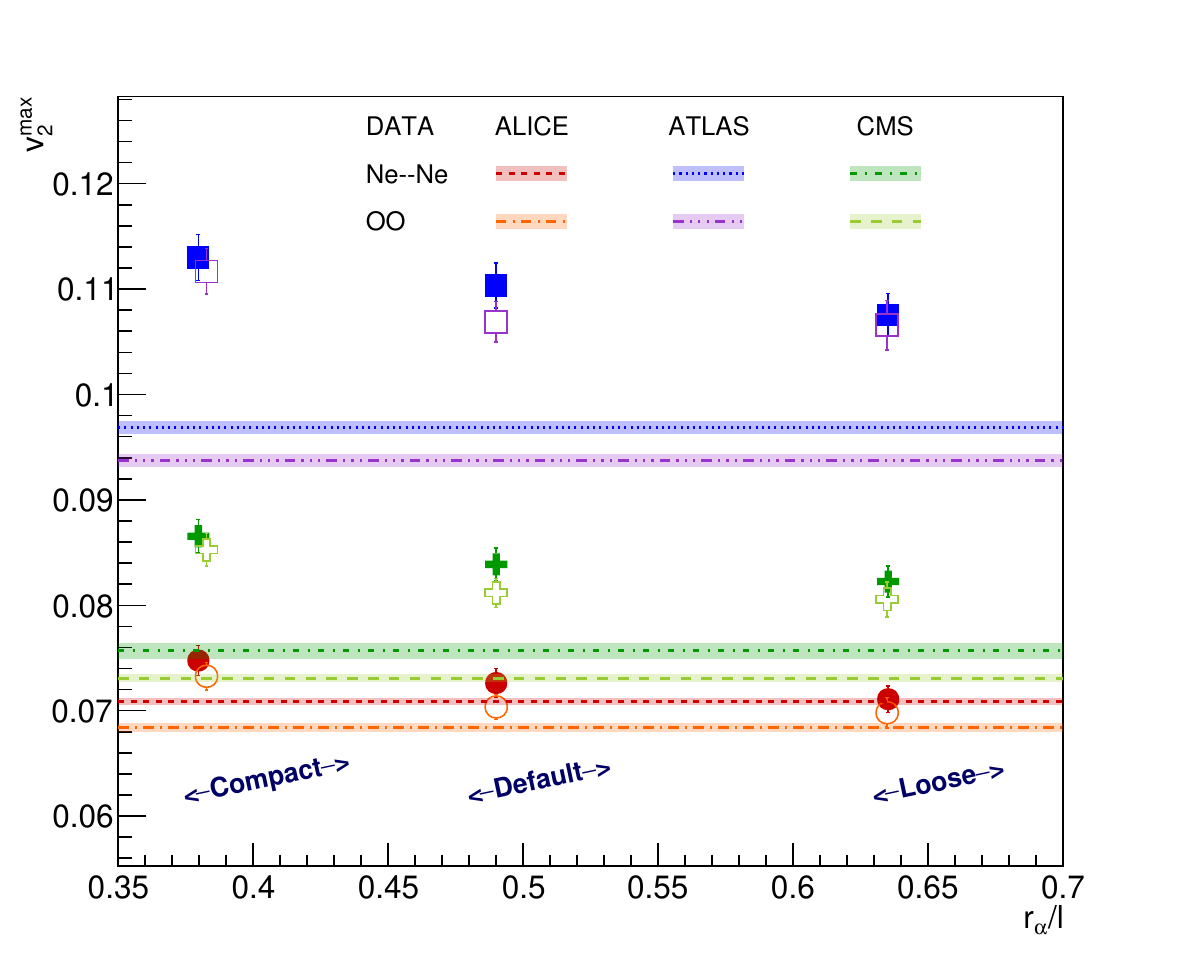}
\includegraphics[scale=0.42]{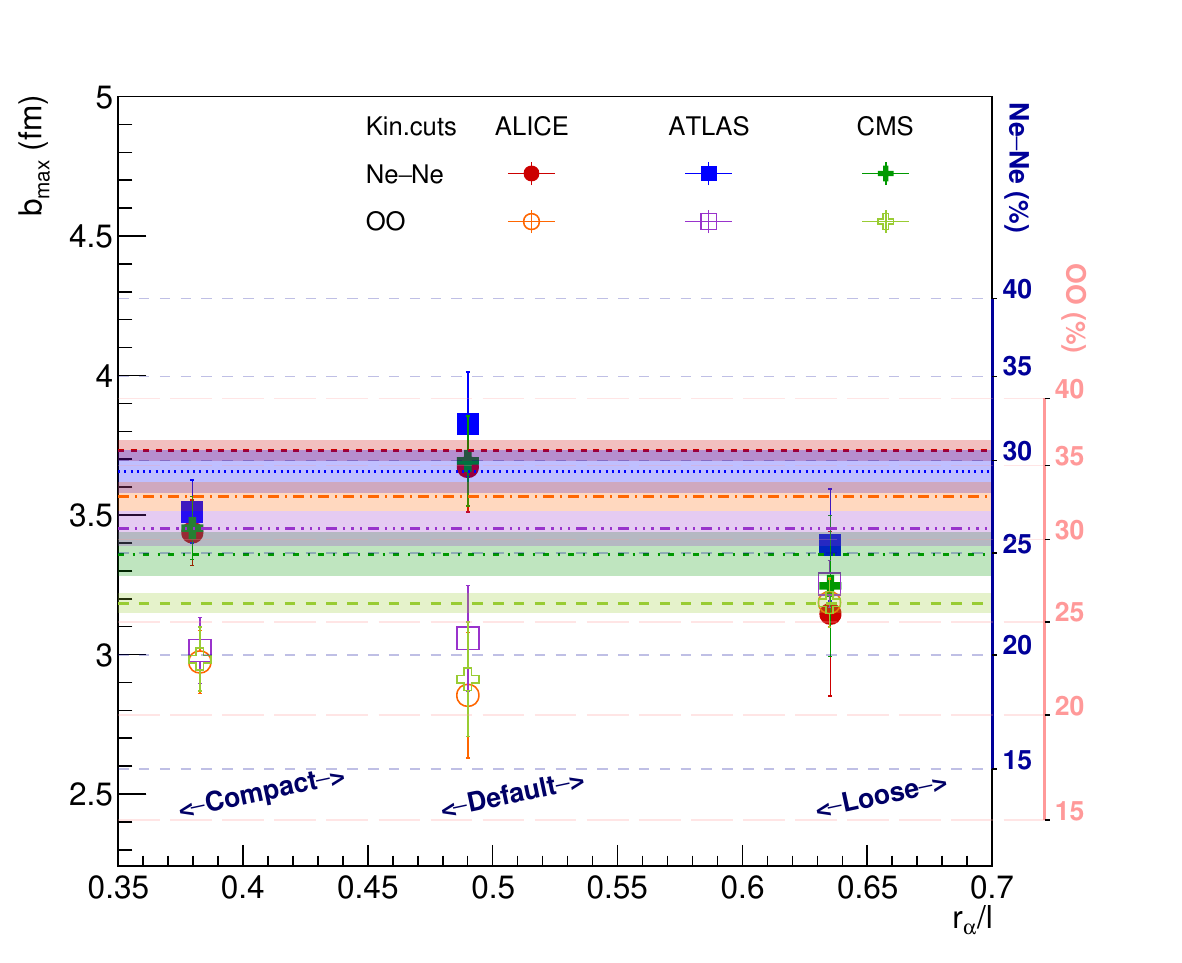}
\caption{Cluster parameter dependence of $v_{2}^{\rm max}$ (extracted from Gaussian fits of $v_{2}$ versus centrality trend) and the impact parameter, $b_{\rm max}$ correlated to corresponding collision centrality, $C_{\rm max}$ for compact, default and loose $\alpha$-clustered nuclear configurations in OO and Ne--Ne collisions at $\sqrt{s_{\rm NN}}~=~5.36$~TeV. Horizontal bars correspond to extraction from experimental data, while markers denote IP-Glasma+MUSIC+iSS+UrQMD model results.}
\label{fig:v2maxBmax}
\end{figure*}

In fact, the peak value of $v_{2}\{ 2 \}$ (denoted as $v_{2}^{\rm max}$) naturally encodes the competing influence of initial nuclear structure and the impact parameter-driven collision geometry on the hydrodynamic evolution of the system. Given the system sizes expected in both OO and Ne--Ne collisions are comparable, any differences in the trends of $v_{2}$ and $v_{2}^{\rm max}$ can be considered originating from the differences in the initial nuclear density profiles, which can modify the $\epsilon_{2}$ in distinct ways~\cite{MenonKavumpadikkalRadhakrishnan:2025apq}. Therefore, to investigate the sensitivity of $v_{2}^{\rm max}$ in probing the underlying nuclear structure and to specifically understand the effect of the compactness of $\alpha$-cluster on the elliptic flow coefficient $v_{2}$, $v_{2}^{\rm max}$ and the impact parameter values corresponding to those centrality classes (\textit{i.e.} $b_{\rm max}$) are plotted with respect to $r_{\alpha}$ (upper row), the inter-cluster distance $l$ (middle row) and $r_{\alpha}$/$l$ ratio for the compact, default and loose $\alpha$-cluster cases of OO and Ne--Ne collisions in Fig.~\ref{fig:v2maxBmax}. To obtain the $v_{2}^{\rm max}$, the $v_{2}$ vs. centrality curve from every $\alpha$-clustered profile, as well as from the ALICE, CMS, and ATLAS experimental results, is fitted with a Gaussian distribution, while the $b$-to-centrality correlation is derived using a third-order polynomial fit. The horizontal bands represent the values corresponding to experimental results. As explained in previous sections, an increasing $r_{\alpha}$(decreasing $l$) implies moving from compact to loose $\alpha$-cluster configurations. It can be seen in Fig.~\ref{fig:v2maxBmax} that the $v_{2}^{\rm max}$ values are slightly higher for Ne--Ne collisions than OO collisions and they drop as function of $r_{\alpha}$ and $r_{\alpha}$/$l$ (or increase as function of $l$), as one moves from compact to default and loose $\alpha$-cluster cases. However, this decrement of $v_{2}^{\rm max}$ for loose $\alpha$-cluster configuration with respect to compact configuration in OO(Ne--Ne) collisions is only by $\approx$ 3.6$\%$(4.4$\%$) in ATLAS, 4.7$\%$(5.8$\%$) in CMS, and 5.5$\%$(5.3$\%$) in ALICE model cases. It is noteworthy that the Gaussian-fit results for $v_{2}^{\rm max}$ from the model and experimental data are once again in good agreement for the ALICE kinematic range of study, for both OO and Ne--Ne collisions. 

\begin{figure*}
\centering
\includegraphics[scale=0.29]{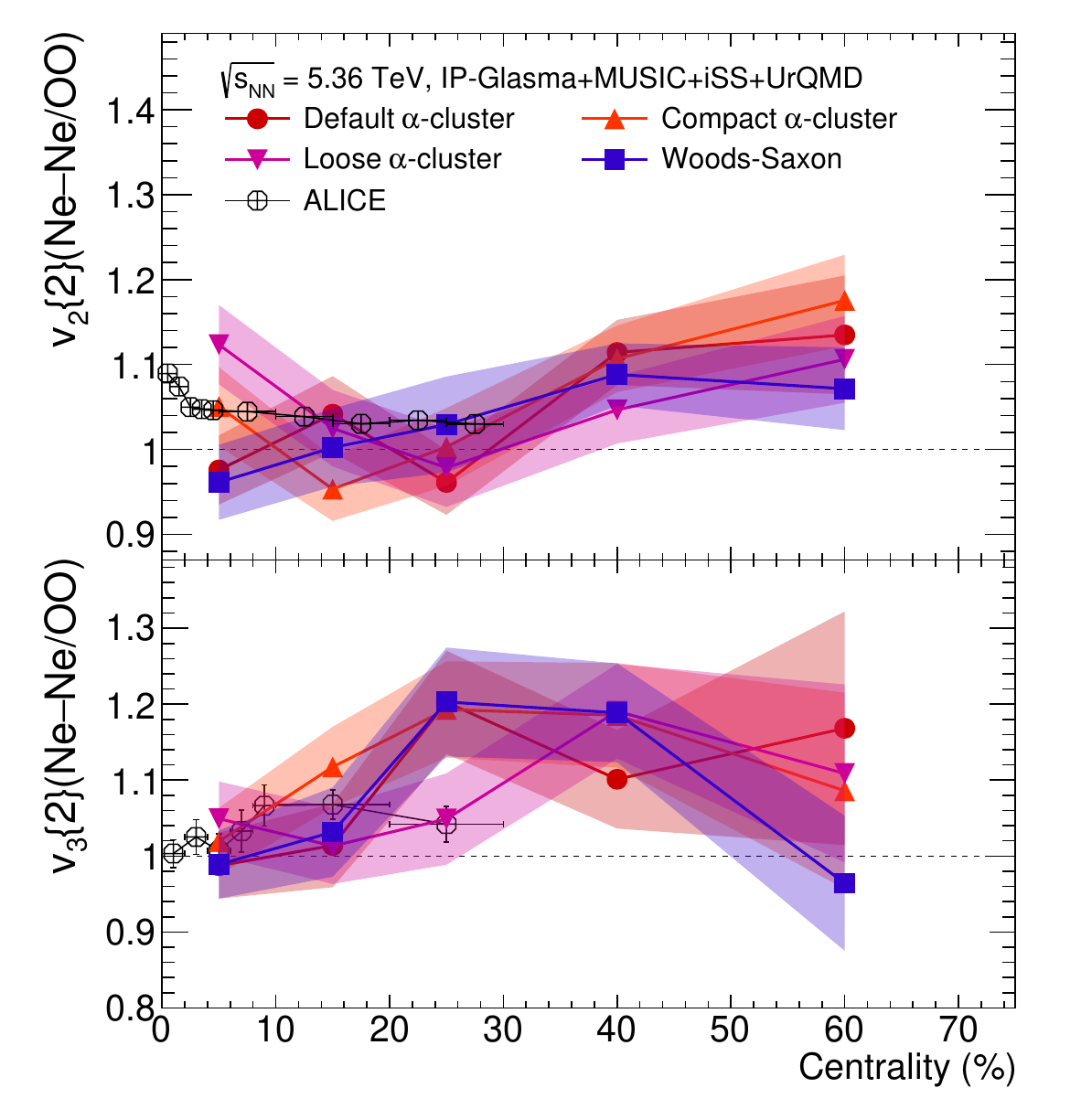}
\includegraphics[scale=0.29]{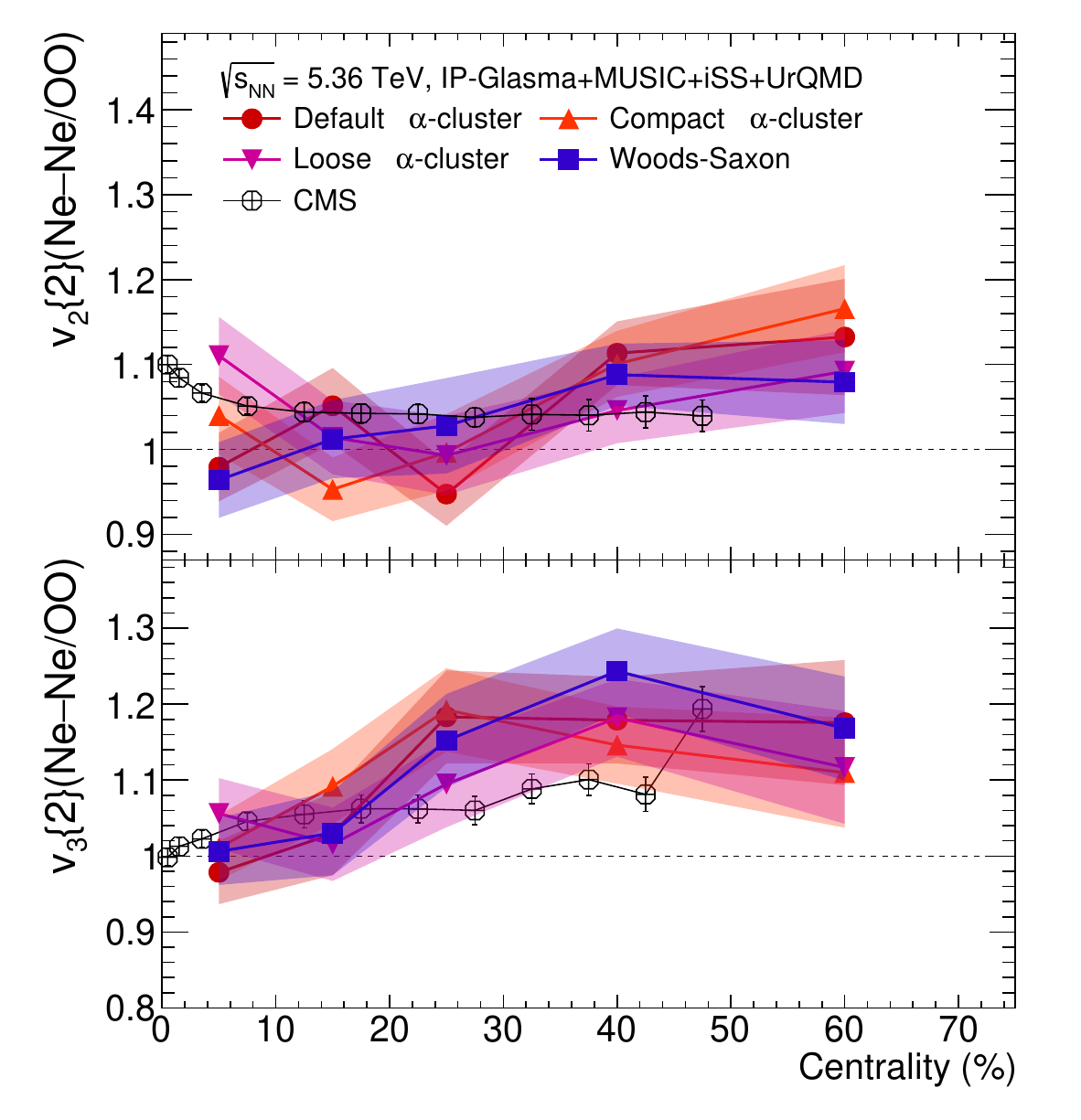}
\includegraphics[scale=0.29]{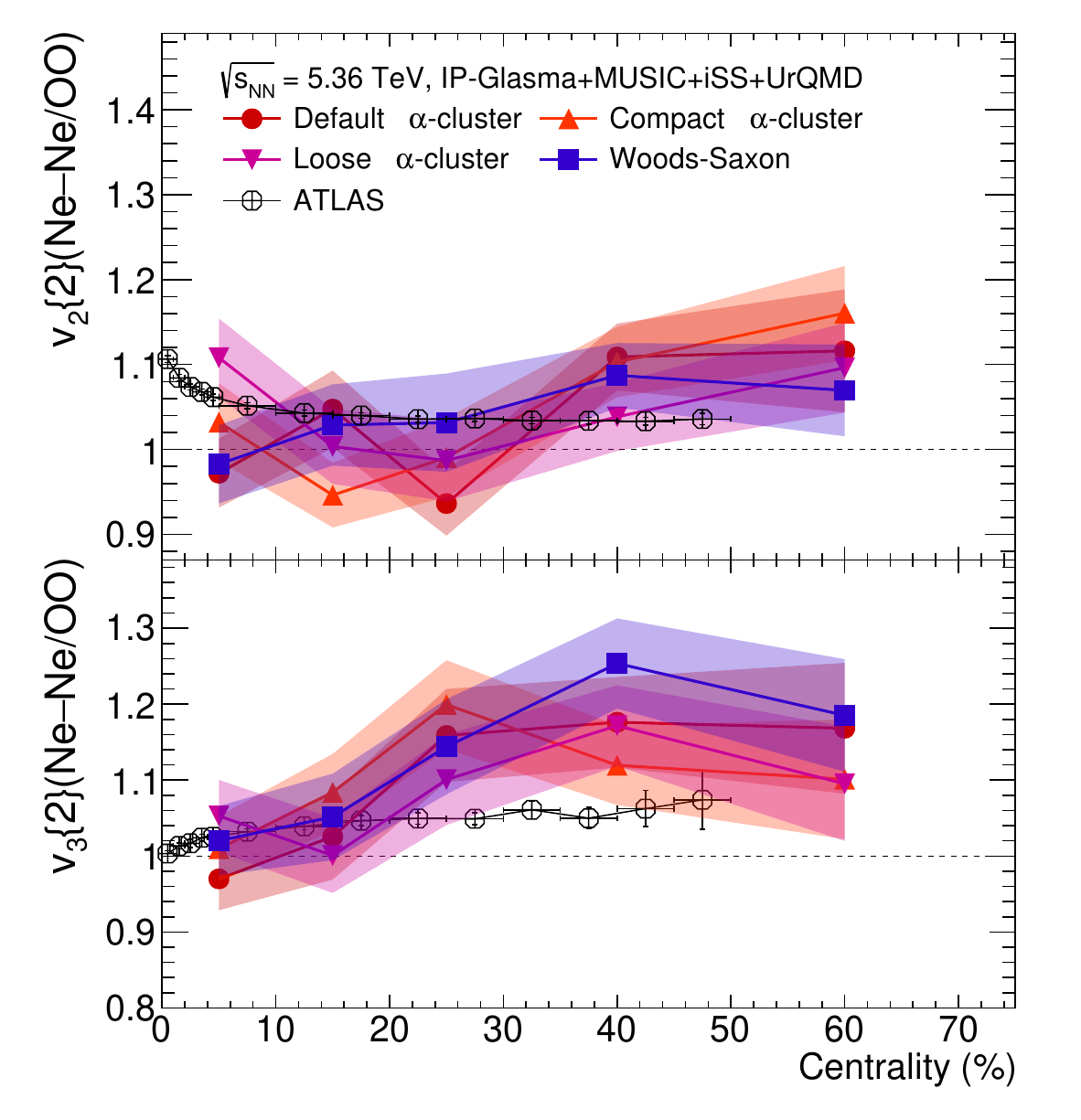}
\caption{Ratios $v_{2}\rm (Ne-Ne/OO)$ and $v_{3}\rm (Ne-Ne/OO)$ as a function of centrality for Woods-Saxon and compact, default, and loose $\alpha$-clustered nuclear configurations in ALICE, CMS and ATLAS kinematic regimes of study using IP-Glasma+MUSIC+iSS+UrQMD model, along with comparison to corresponding experimental data~\cite{ALICE:2025luc,ATLAS:2025nnt,CMS:2025tga}. Error bands incorporate the statistical uncertainties from model estimations.}
\label{fig:NeORatio}
\end{figure*}

To identify the most favourable collision geometry that generates elliptic flow results qualitatively and quantitatively closer to OO and Ne--Ne experimental results and to get additional hints on the degree of localization of $\alpha$-clusters, $b_{\rm max}$ is studied as a function of clustering parameters and is compared to those estimated from experimental data as already described earlier. It is observed that the $v_{2}^{\rm max}$ from Gaussian-fits is attained in the (20--25$\%$), (17--27$\%$) and (25--30$\%$) centrality classes respectively for compact, default and loose $\alpha$-cluster cases of OO collisions while it is in the (23--29$\%$), (27--35$\%$) and (17--29$\%$) centrality classes respectively for Ne--Ne collisions. The decreasing cluster compactness in OO collisions is found to result in a gradual shift of $b_{\rm max}$ to higher values, reinforcing how central collisions are better in revealing the tightly bound $\alpha$-clusters in a tetrahedral $^{16}\rm O$ nucleus~\cite{Constantin:2025ova, Ding:2023ibq, MenonKavumpadikkalRadhakrishnan:2025apq, Behera:2023nwj, R:2024eni, Li:2020vrg}. However, the bowling-pin structure of $\alpha$-clustered $^{20}\rm Ne$ nuclei contributes differently to the development of $v_{2}$ and therefore $v_{2}^{\rm max}$; $v_{2}^{\rm max}$ is achieved at the earliest impact parameters by the loose and then the compact $\alpha$-clustered Ne--Ne collisions in comparison to the default case. As far as the agreement of the Gaussian-fit results of both model calculations and experimental data is concerned, the loosely clustered configuration is in better agreement for both OO and Ne--Ne collisions, while the compact clustered profile also has a fair overlap in the case of Ne--Ne collisions. This is equally evident from $b_{\rm max}$ versus $r_{\alpha}$/$l$ plot also in Fig.~\ref{fig:v2maxBmax}. In brief, the Gaussian fits suggest that the highest $v_{2}$ is reached by the compact $\alpha$-clustered configuration, followed by the default, and then the loose case for both collision systems. Additionally, the attainment of $v_{2}^{\rm max}$ keeps shifting to higher $b$ values as we increase the $r_{\alpha}$/$l$ ratio for OO collisions as well as for the compact and default $\alpha$-clustered Ne--Ne collisions, owing to the uniform reduction in the clustering effect in the Oxygen and Neon nuclei as visualized in Fig.~\ref{fig:ONe3config}. However, the interpretation of the loosely clustered Ne--Ne collisions is as follows: the most central Ne--Ne collisions with their loose configuration provide one of the best elliptic overlap due to the very 5-clustered Neon nuclear structure and hence a significant $v_{2}$, while a second $v_{2}^{\rm max}$ is obtained around 40$\%$ centrality class owing to the geometry of the overlap itself (see Fig.~\ref{fig:NeO-v23-3expmnts}). A Gaussian fit would average out these double peaks, providing the result seen in Fig.~\ref{fig:v2maxBmax}. It is interesting to note the similarity in $b_{\rm max}$ between the Gaussian fit of the real experimental data represented by the horizontal bands and the hybrid hydrodynamic model prediction.


One of the robust methods suggested, which can isolate the nuclear structure differences of two light ions with comparable mass numbers (here, $^{16}\rm O$ and $^{20}\rm Ne$), is to study the ratio of their flow properties. As the systems produced after their collision are expected to undergo similar hydrodynamic evolution, the ratios such as $v_{\rm n}$(Ne--Ne/OO) can effectively reduce the common final-state collective effects, thereby highlighting the relevant differences arising from intrinsic nuclear geometry like deformations, $\alpha$-clustering, etc., and the initial-state fluctuations~\cite{ALICE:2025luc,ATLAS:2025nnt,CMS:2025tga}. Figure~\ref{fig:NeORatio}, therefore presents, $v_{\rm 2}$(Ne--Ne/OO) and $v_{\rm 3}$(Ne--Ne/OO) as a function of centrality for the four nuclear configurations, for the ALICE, CMS, and ATLAS regimes of study, with a comparison to corresponding data points available from the three respective experiments. The ratio peaks observed for $v_{2}$ in the ultra-central collisions by all three experiments, attributed to the stronger quadrupole deformation of $^{20}\rm Ne$ relative to tetrahedral $^{16}\rm O$ are qualitatively mimicked by the loose and compact $\alpha$-cluster configurations, while a better agreement with the rest of the centrality classes are observed only for the loose cluster case. On the other hand, the smaller $v_{3}$ ratios in central collisions, attributed to the larger octupole deformation of tetrahedral $^{16}\rm O$, and the gradual increase of the ratio with increasing centrality, are not well-reproduced by any single nuclear configuration. In fact, beyond 20$\%$ centrality, the model predictions overestimate $v_{\rm 3}$(Ne--Ne/OO), though the increasing trend is reproduced by all configurations considered in this study.

\section{Summary}
Using a hybrid hydrodynamic model, this work performs a systematic study of $v_{2}\{2\}$ in OO and Ne--Ne collisions at $\sqrt{s_{\rm NN}}~=~5.36$~TeV for varying degrees of $\alpha$-clustering inside the colliding nuclei, namely, compact, default, and loose cluster configurations, in contrast to the Woods-Saxon nuclear profile. Elliptic flow is estimated for three kinematic regions--- corresponding to the ALICE, CMS, and ATLAS experimental acceptances. We observe that the impact parameter-based centrality classification reveals a clear dependence of $v_{2}$ on collision centrality and the initial nuclear configuration, particularly in OO collisions. As the clustered configuration changes from compact to loose, the magnitude of $v_{2}$ peaks is seen to drop as well as shift to higher $b$ values---hence supporting our understanding that the initial nuclear profile has a significant role in the development of final-state elliptic flow. The one configuration whose $v_{2}^{\rm max}$, $b_{\rm max}$, $v_{2}\rm (Ne-Ne/OO)$ and $v_{3}\rm (Ne-Ne/OO)$ consistently match the experimental data qualitatively (and quantitatively in the case of ALICE), is the loose $\alpha$-cluster configuration of $^{16}\rm O$ and $^{20}\rm Ne$ nuclei. This does not necessarily imply that the loose $\alpha$-cluster configuration represents the true ground-state structure of the $^{16}$O and $^{20}$Ne nuclei. Rather, it suggests that the measured flow observables are better described by a diffuse nuclear density profile closer to the Woods--Saxon distribution than by a strongly localized $\alpha$-cluster configuration. Further, this may indicate that, if $\alpha$-clustering is present, it is likely to be less pronounced than assumed in the compact cluster models, where clustered configurations contribute only partially to the nuclear wave function rather than dominating it.

\section{Outlook}
The recent light-ion collision program at the LHC has opened a unique opportunity to probe the intrinsic structure of light nuclei through collective flow measurements. The present framework can be readily extended to other flow observables, such as higher-order flow harmonics, flow fluctuations, and symmetric cumulants, offering a comprehensive approach to constraining nuclear deformation and $\alpha$-clustering in light nuclei. Machine learning techniques trained on realistic event-by-event simulations can be used to estimate the collision geometry, especially the impact parameter, from final-state observables~\cite{Mallick:2021wop}. This will help reduce the biases associated with multiplicity-based centrality selection and allow more precise constraints on the nuclear structure of light nuclei using relativistic heavy-ion collisions. Therefore, the methodology presented in this work, together with data-driven approaches for geometry-based centrality estimation, can be applied in future experimental analyses. Precise measurements of anisotropic flow coefficients, combined with improved reconstruction of the collision geometry, will provide valuable insights into the $\alpha$-cluster structure of light nuclei and its influence on the development of final-state collectivity in light-ion collisions.

\section*{Acknowledgement}
AMKR acknowledges the doctoral fellowship from the DST INSPIRE program of the Government of India. GGB and SP gratefully acknowledge the Hungarian National Research, Development and Innovation Office (NKFIH) under Contract No. NKFIH NEMZ\_KI-2022-00058, 2024-1.2.9-NETWORKING-2024-00047, 2024-1.2.5-TET-2024-00022, and Wigner Scientific Computing Laboratory (WSCLAB, the former Wigner GPU Laboratory). NM is supported by the Academy of Finland through the Center of Excellence in Quark Matter with Grant No. 346328. RS sincerely acknowledges the DAE-DST, Government of India, funding under the mega-science project – “Indian participation in the ALICE experiment at CERN” bearing Project No. SR/MF/PS-02/2021-IITI (E-37123). The authors gratefully acknowledge the MoU between IIT Indore and Wigner Research Centre for Physics (WRCP), Hungary, for the techno-scientific international cooperation.


\end{document}